\documentclass[final,1p,times]{elsarticle}

\usepackage{amssymb}
\usepackage{lscape} 
\usepackage{longtable}
\usepackage{tabularx}
\usepackage{caption}
\usepackage{array}
\usepackage{multirow}
\usepackage{ltxtable}
\usepackage{url}
\usepackage{multirow}
\usepackage{multicol}
\usepackage{supertabular}

\usepackage{nomencl}
\makenomenclature

\nomenclature{TTP}{Trusted Third Party}
\nomenclature{RNE}{Road Network Embedding}
\nomenclature{BT}{Blockchain Technology}
\nomenclature{MaaS}{Mobility-as-a-service }
\nomenclature{CAGR}{Compound Annual Growth Rate}
\nomenclature{ICT}{Information and communications technology}
\nomenclature{LBS}{Location-based services}

\journal{}

\begin{document}

\begin{frontmatter}



\title{Data privacy for Mobility as a Service}

\author[1,4]{Zineb Garroussi\corref{cor1}}
\ead{zineb.garroussi@polymtl.ca}

\author[1,2,4]{Antoine Legrain}
\ead{antoine.legrain@polymtl.ca}

\author[3]{Sébastien Gambs}
\ead{gambs.sebastien@uqam.ca}

\author[5]{Vincent Gautrais}
\ead{vincent.gautrais@umontreal.ca}

\author[1,4]{Brunilde Sans\`o}
\ead{brunilde.sanso@polymtl.ca}

\cortext[cor1]{Corresponding author}

\affiliation[1]{organization={Polytechnique Montreal},
addressline={2900 Edouard-Montpetit Blvd, University of Montreal Campus, 2500 Polytechnique Road},
postcode={H3T 1J4},
city={ Montreal},
country={Canada}}

\affiliation[2]{organization={Interuniversity Research Center on Enterprise Networks, Logistics and Transportation (CIRRELT)},
addressline={Andre-Aisenstadt Building, University of Montreal, P.O. Box 6128},
postcode={H3C 3J7},
city={ Montreal},
country={Canada}}

\affiliation[3]{organization={University of Quebec in Montreal, Computer science department},
addressline={2098 Rue Kimberley},
postcode={H3C 3P8},
city={ Montreal},
country={Canada}}

\affiliation[4]{organization={Group for Research in Decision Analysis (GERAD)},
addressline={HEC Montreal, 3000, Cote-Sainte-Catherine Road},
postcode={H3T 2A7},
city={ Montreal},
country={Canada}}

\affiliation[5]{organization={CRDP - Faculty of Law - University of Montreal},
addressline={3101 Ch de la Tour},
postcode={H3T 1J7},
city={ Montreal},
country={Canada}}


\begin{abstract}
Mobility as a Service (MaaS) is revolutionizing the transportation industry by offering convenient, efficient and integrated transportation solutions. 
However, the extensive use of user data as well as the integration of multiple service providers raises significant privacy concerns. 
The objective of this survey paper is to provide a comprehensive analysis of the current state of data privacy in MaaS, in particular by discussing the associated challenges, existing solutions as well as potential future directions to ensure user privacy while maintaining the benefits of MaaS systems for society.
\end{abstract}

%

\begin{keyword}
Data privacy \sep Location-based Services \sep Mobility as a Service (MaaS) \sep  Privacy by design  \sep Ride-hailing \sep Ridesharing.


\end{keyword}

\end{frontmatter}
\section{Introduction}
\label{intro}

The growing urbanization of our society, along with the increasing availability of mobile technology, is driving the need for alternative mobility solutions. 
For instance, according to the United Nations, 55\% of the world's population now live in urban areas, and this number is expected to grow to 68\% by 2050~\cite{unworld18}.
Overall combined with population growth, another 2.5 billion people are thus expected to join cities over the next 30 years.
Among possibles approaches, we focus on mobility-as-a-service (MaaS) solutions in this paper.  

More precisely, MaaS uses information and communication technologies to integrate real-time traffic data with existing transportation modes, such as public transport, cars, bikes and rideshares. Thus, MaaS enables optimal route display and one-stop registration, reservations and payments through apps like Apple Pay or credit cards~\cite{reyes20}, thus enhancing convenience. 

Originating in Finland in 2016, MaaS uses advanced technologies like IoT to optimize transportation~\cite{butler21,kamargianni16}.
Since then, MaaS has revolutionized the transportation system in a way that no one could have even imagined a decade ago. 

According to Statistica, the global MaaS market is estimated to be worth over \$230 billion by the end of 2025 (at a CAGR of 32.6\%)~\cite{statista2023}. 
Its success is also measured by factors like adoption rate and environmental impact, showing its ability to encourage sustainable travel and economic growth~\cite{pandey19}. 
In particular, MaaS helps to reduce private car usage, mitigates climate change, alleviates traffic congestion and facilitates multi-modal travel with ease. 
It can also foster people-focused communities, boost local industries and aids tourism by improving access to remote areas~\cite{sunghi21, benjamin22}.

However, the rapid expansion of MaaS has led to an increasing reliance on personal data to deliver seamless and efficient transportation solutions~\cite{pagoni22}. 
With this growth comes a pressing need to address privacy concerns and implement recognized privacy practices, such as privacy-by-design, privacy-by-default as well as privacy impact assessments~\cite{cottrill20}. 
In particular, the right to privacy is now viewed as a fundamental human right in many national constitutions, international treaties, and the Universal Declaration of Human Rights, and is gaining increased prominence in the modern data-driven landscape~\cite{kuner20,ohchr21}. 
Key privacy regulations, like the EU's General Data Protection Regulation, provide individuals control over their personal information, shielding them from unwarranted intrusions~\cite{tikkinen18}. 
Yet, the widespread utilization of artificial intelligence and data mining for handling massive personal data quantities is kindling new concerns~\cite{timan21}.

These concerns are tied to the growing data generation and collection by governments and corporations, and the subsequent questions about ethical data use and access. 
Consequently, there are calls for robust privacy protections mechanisms and regulations, to ensure individuals' control over their personal information.
Privacy by Design, a concept introduced by Ann Cavoukian in 1995, advocates for proactive embedding of privacy into technology systems, rather than as a later addition~\cite{cavoukian10}. This principle promotes seamless and transparent user privacy, requiring no additional user action~\cite{alnajrani20}. Nowadays acknowledged worldwide, it is recognized as a best practice in personal data protection in today's digital world \cite{friginal14}. 

The approach's seven principles~\cite{cavoukian10,cavoukian09}, which include for instance proactive privacy, privacy as default and user respect, aim to anticipate privacy issues, maintain system usability as well as ensuring user control over their personal data. 

In particular, this proactive privacy incorporation guides our discussion on data privacy within the evolving MaaS ecosystem.
To realize this, privacy-enhancing technologies have emerged as part of the solution by aiming to reduce personal data collection, processing and sharing. 
These technologies help to implement privacy-by-design principles~\cite{schaar10}, considering privacy at every design stage.
Examples of privacy-enhancing technologies include techniques such as encryption and anonymization~\cite{cottrill20,karagiannis20}. 

In the context of privacy protection, the concept of privacy-by-default is crucial. 
It is enshrined in the EU's General Data Protection Regulation (GDPR) and is aligned with privacy-by-design. 
According to Article 25 of the GDPR, controllers are required to process only the necessary personal data, ensuring limited and proportionate data processing while respecting individuals' privacy rights. 
The GDPR also obliges controllers to implement measures for data security, safeguarding against unauthorized processing, accidental loss, destruction or damage.

Finally, Privacy Impact Assessments (PIAs) play a significant role in evaluating the potential impact of projects, policies or systems on individuals' privacy rights~\cite{clarke09}. 
These assessments, recommended by the Office of the Privacy Commissioner of Canada (OPC), identify privacy risks and propose appropriate mitigation measures. 
Although not always legally mandated in Canada, PIAs are considered best practices for demonstrating compliance with data protection laws and some Canadian provinces like British Columbia require their implementation \cite{georgiadis22}.

Focusing on a specific type of risk, as the popularity of MaaS applications increases, there's a surge in the collection, storage, sharing and tracking of real-time, location-based data.
More precisely, in the MaaS ecosystem, users engage with providers via smartphones, which enables continuous tracking of location and behavior for profiling and other purposes~\cite{huang22}.
This collection includes for example users' travel patterns, destinations and modes of transportation. 
While this information can be utilized for various purposes such as marketing and advertising, it may intrude upon users' privacy~\cite{paiva20}. 
In addition, location data, considered to be sensitive by nature, can reveal personal information such as religious beliefs or political affiliations, hence requiring heightened protection. 
EU data protection law recognizes these risks and demands specific protections for special categories of personal data, which may pose high risks when processed. 
While, MaaS providers collect location data to personalize subscriptions and analyze travel patterns, the danger lies in the potential for unrestricted access leading to the unfair accumulation of movement profiles, and the ``datification'' of patterns for the purpose of shaping and selling personalized goods and services \cite{panahi22}.

In particular, the widespread use of technology and Internet has facilitated data collection and storage for organizations, but it has also elevated the risk of data breaches and unauthorized access to sensitive information~\cite{zheng18, huang22}.

MaaS applications, given their ability to collect real-time, detailed data, significantly elevate the privacy risks traditionally associated with data collection. MaaS systems are susceptible to various attacks that threaten personal data, re-identification, inference and eavesdropping attacks. Additionally, malware can potentially access a user's device and steal data, while social engineering attacks deceive users into sharing information. Thus, safeguarding against these attacks is vital for preserving the privacy and security of personal data in MaaS systems~\cite{callegati18}.
\paragraph{Scope of the Paper} This paper aims to present a comprehensive survey of the current state of data privacy within the MaaS ecosystem. More specifically, the scope of this survey encompasses several key areas, including the identification of major privacy challenges in MaaS, such as risks related to personal information, location data and third-party data sharing. 
Furthermore, it examines existing privacy-enhancing techniques employed in MaaS systems, such as anonymization methods, cryptographic solutions, blockchain-based approaches and privacy-preserving machine learning. Additionally, the paper delves into the legal and regulatory landscape governing data privacy in MaaS, with a focus on international privacy standards and legislation. Finally, this survey offers recommendations for enhancing data privacy in MaaS and identifies potential future research directions in this critical field.
\paragraph{Methodology} To gather relevant papers on the subject of data privacy in the MaaS ecosystem, a systematic and comprehensive approach was adopted. Multiple academic databases and search engines, such as IEEE Xplore, Google Scholar and ScienceDirect were utilized to identify  papers that discussed the various aspects of data privacy within MaaS. 
Key search terms and phrases, including "Mobility as a Service", "data privacy", "privacy challenges", "privacy-preserving techniques" and "legal and regulatory in MaaS", were employed to ensure the thoroughness of the search process. 
The collected literature was then carefully scrutinized for quality and relevance, with an emphasis on studies that provided insights into the current state of data privacy in MaaS, existing privacy-enhancing techniques as well the regulatory environment surrounding this domain.
\paragraph{Paper Organization and Structure}
The remainder of the paper is structured as follows. 
First, Section~\ref{data_in_maas} delves into the state of data in MaaS, with an emphasis on location-based services data and related collection and processing methods. 
Afterwards, Section~\ref{privacy_laws_and_regulations_in_maaS} explores the legal and regulatory framework for data privacy in MaaS, addressing local and international data protection laws and compliance challenges. 
In Section~\ref{classification_of_data_privacy_methods_in_maas}, the classification of data privacy methods in MaaS is presented, discussing related works and identifying limitations in the existing state-of-the-art. 
Then, Section~\ref{future_research_directions} highlights future research directions in data privacy for MaaS, such as game theory, edge computing and evaluation methods. 
Finally, Section~\ref{conclusions}  concludes the paper.
\section{Data in MaaS}
\label{data_in_maas}
Data, consisting of stored electronic information, symbols, or signals, is essential for the success of Mobility as a Service (MaaS)~\cite{aman22}. In particular, MaaS providers amalgamate a vast array of transport data and APIs from multiple operators to allow customers access to information on mobility options, reservation and payment. 
APIs manage interactions between web-based services, constituting the foundation for MaaS providers' operations \cite{itf2023}. 
Through API sharing, MaaS providers present a unified digital platform for planning, booking, paying and utilizing transportation while also creating the app and website that function as the main customer-facing interface~\cite{haoning22}. MaaS applications manage directly several categories of sensitive data, which include but are not limited to Personal Identifiable Information (PII) and location data.

\subsection{Personal Identifiable Information}

MaaS platforms gather a broad spectrum of PII from their users, including but not limited to names, physical addresses, email addresses and phone numbers, which are typically required for account creation. 
Other identifying information such as IP addresses and device identifiers are also often collected. 
Furthermore, payment data like credit or debit card details and billing addresses are necessary for transactional purposes.
In case of a breach, such data is susceptible to lead to identity theft, financial fraud as well as other forms of cybercrime.
Additionally, MaaS platforms may collect travel history, such as the details of journeys taken, preferred routes, frequency of travel and times of travel. 
This kind of PII can be used to tailor services, suggest personalized routes and enhance overall user experience. 
User preference data, such as preferred modes of transport or payment methods, can also be recorded to facilitate ease of use and personalized recommendation provided to a user.
Additionnally, biometric data such as facial recognition or fingerprint information, used for identity verification and access control, can also be sensitive as it can be used for surveillance or other forms of privacy intrusion.
Finally, demographic information like age, gender and occupation may be obtained to aid in broad-level analysis and segmentation~\cite{cottrill20}.

\subsection{Location data}

Location-based services (LBS) in MaaS refers to the use of location to provide users with personalized and context-aware mobility options and services based on their current location and travel preferences. 
LBS can include features such as real-time transit updates, trip planning and optimization, location-based promotions and discounts and on-demand ride-hailing services~\cite{liu18}. 

For instance, a MaaS platform might employ LBS to provide the most practical and economical travel options depending on a user's present location, the time of day, and preferred mode of transportation. This could include recommending public transit options with real-time updates on delays or suggesting an on-demand ride-hailing service for last-mile connectivity~\cite{lin17}. However, the use of LBS in MaaS presents privacy and security concerns because it involves the gathering and use of location data. Thus, appropriate safeguards and user consent mechanisms should be in place to ensure the responsible use of location data in MaaS services. 
In particular, location data can reveal sensitive information about users in several ways. First, it can correspond to real-time location information, as well as location history, which can expose users' movements, habits and routines, potentially leading to stalking, harassment or other malicious activities \cite{jiang21}. 
Second, location data can reveal health-related information, such as medical conditions, medication regimes and mobility limitations, which can be exploited for discrimination, targeted advertising or other harmful purposes \cite{zhang22}. 

Privacy is a key concern in both continuous and snapshot LBS, as these services rely on the collection and use of sensitive location data. In continuous LBS, user location is tracked continuously over time, which can lead to more severe privacy risks. It can be used to create detailed movement profiles of individuals, which can reveal sensitive information about their habits and routines. This information could be misused for stalking, surveillance and other malicious purposes~\cite{shaobo19}. 
To reduce the privacy exposure in continuous LBS, authors in~\cite{shaobo19} proposed a privacy method based on multi-level caching mechanism and spatial $k$-anonymity. A detailed discussion on k-anonymity will be presented in the section \ref{classification_of_data_privacy_methods_in_maas}. In snapshot LBS, user location data is obtained at a specific point in time, which may seem less intrusive than continuous tracking.
However, snapshot LBS can still pose privacy risks if the position of a user is collected and used without appropriate safeguards and user consent. 
For example, location-based advertising and recommendations may use user location data to provide personalized offers or promotions, but users may not be aware of how their location data is being used or shared with third-party advertisers.

To address these privacy concerns, it is important to implement appropriate privacy and security measures in both continuous and snapshot LBS, such as data anonymization, data minimization and user consent mechanisms. 
\cite{hwang14}~aims to address the issue of privacy leakage from location data by obfuscating the temporal information in trajectory data.

The algorithm uses a randomized time transformation method to create a time-obfuscated trajectory that preserves the spatio-temporal properties of the original trajectory while hiding the exact time information. 
The authors evaluated the proposed algorithm using real-world datasets and compared it to existing trajectory privacy protection methods.
The results showed that the proposed algorithm achieved better privacy protection performance while maintaining a high level of data utility. 
In~\cite{dua21} a proposed mechanism utilizes a deep learning-based approach to extract the location features from the original location data, which is then used to generate a privacy-preserving representation of the location data. This privacy-aware representation ensures that direct identification from the location data becomes unfeasible, thereby safeguarding user privacy. Importantly, despite this transformation, the resultant representation retains critical characteristics of the original data. This implies that specific tasks, such as location-based service recommendations or mobility pattern analysis, remain feasible with this anonymized data. The strength of this approach lies in its ability to strike a balance between utility and privacy, ensuring that essential features for particular tasks are preserved even as identifiable details are obscured.



 \cite{tao17}~proposes a scheme for preserving the location privacy of users in LBS based on a novel combination of $k$-anonymity and differential privacy techniques to achieve a better trade-off between privacy and utility. The paper discusses the privacy threats in LBS and evaluates the proposed scheme through simulations. \cite{kuang17}~proposes an enhanced privacy-preserving framework for LBS. The proposed framework is based on a double cloaking region scheme, which utilizes two cloaking regions to protect user location information. It should be noted that a comprehensive discussion on differential privacy and cloaking will be presented in section \ref{classification_of_data_privacy_methods_in_maas}.
The framework also incorporates supplementary information constraints to enhance the security and privacy of user data. 
Additionally, regulators and policymakers should establish clear guidelines and regulations around the collection and use of location data in LBS to protect users' privacy rights.

\subsection{The Role of Data Privacy Across Different MaaS Integration Stages}

\begin{figure*}[!h]
\centering
\includegraphics[width=\linewidth]{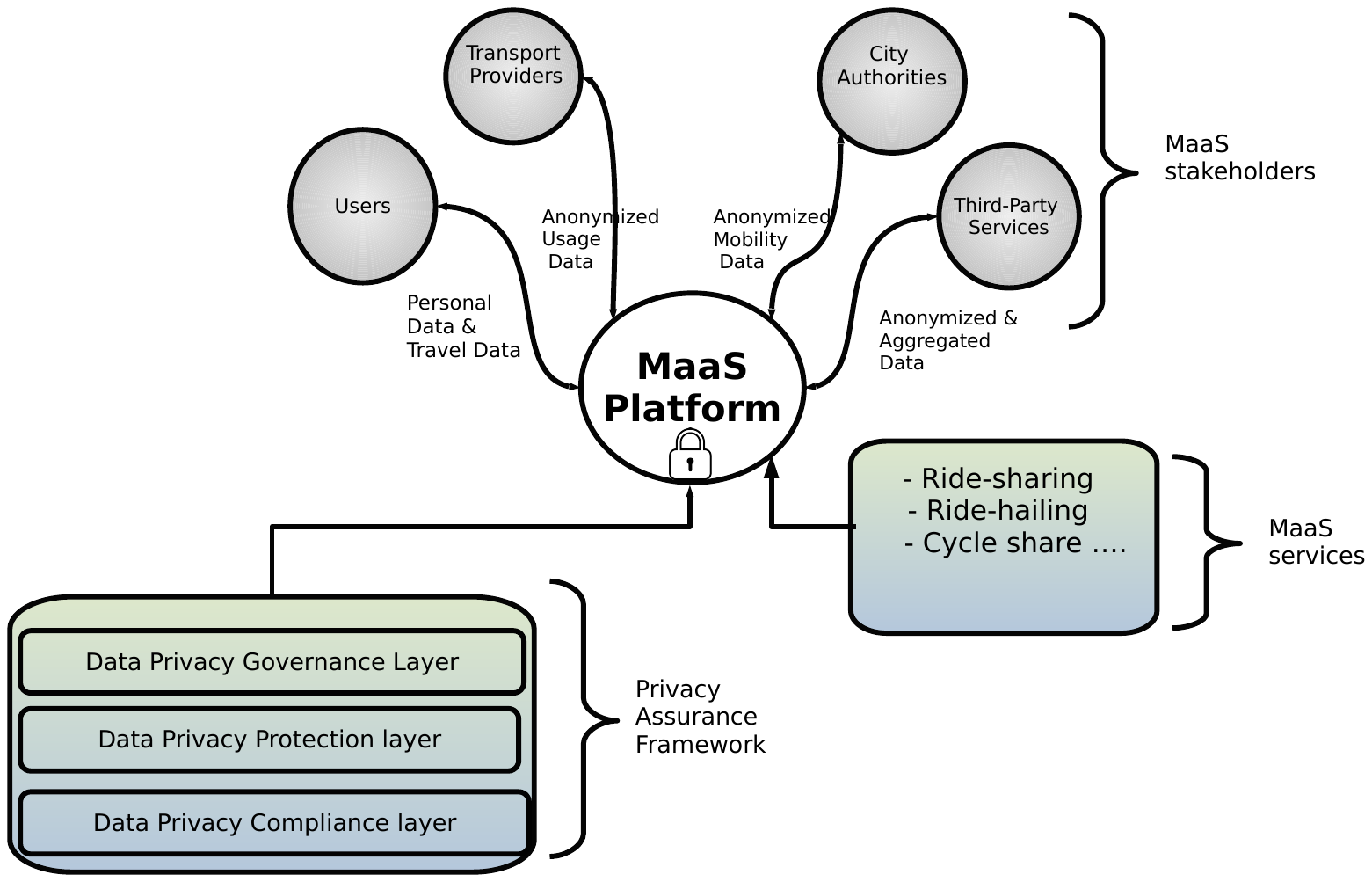}
\caption{MaaS Ecosystem and Privacy Assurance Framework}
\label{maasplatform}
\end{figure*}

As MaaS becomes increasingly complex and integrated, with an evolving ecosystem of stakeholders and services, it becomes critical to manage the growing challenges associated with data privacy.
More precisely, each level of integration in MaaS brings a unique set of data privacy issues. 
Recognizing these challenges, and developing comprehensive strategies to address them, will play a vital role in shaping the future of MaaS~\cite{sohor18}. 
Figure~\ref{maasplatform} provides a visual representation of the MaaS ecosystem, highlighting the different stakeholders involved in data exchange, the services provided by the platform as well as the Privacy Assurance Framework layers that ensure data privacy and protection.

The taxonomy developed by the research team at Chalmers University of Technology in Sweden holds significance in comprehending and tackling data privacy issues within MaaS~\cite{sohor18}. 
As MaaS evolves through its five stages, from Level 0 (No integration) to Level 4 (Policy integration), the degree of data privacy concerns escalates in tandem. 
For instance, a Level 1 (Information Integration) application can function without creating user profiles, thus minimizing data privacy risks. 
However, as MaaS advances to Level 2 (Integration of search/reservation/payment), more personal data is required for tasks such as booking and payment, subsequently increasing privacy concerns. Level 3 (Service Delivery Integration): Further integration of all modes of transportation and movement will be promoted. Regardless of public or private, not only transportation but also sharing services will be added, and all means of transportation will be available. Platforms and subscription services that allow unlimited rides on mobile services within a certain area are also envisioned. It is the integration of service delivery. Level 4 (Policy integration): In addition to efforts at the business level, support from the national and local governments can be obtained. By linking MaaS with policies such as tourism development and smart cities, it is expected to solve social issues such as alleviating traffic congestion and improving convenience for the elderly. 
This classification is crucial for understanding the implications of data privacy as MaaS progresses and becomes more integrated, enabling the development of effective strategies to protect user information while maintaining the convenience and efficiency that MaaS aims to provide.
  
Table~\ref{tab:maas_data_summary} provides a brief summary of the types of data collected from users, transport providers, city authorities and third-party services at each level of the MaaS taxonomy. 
In the table each column is associated to one of the level of the taxonomy, including No Integration Level (Level 0), Level 1 (Information Integration), Level 2 (Integration of Search/Reservation/Payment), Level 3 (Service Delivery Integration) and Level 4 (Policy Integration). 
Each row of the table corresponds to a stakeholder, and the cells provide a summary of the types of data that may be collected from each stakeholder at each level. This table can help MaaS providers understand various types of data that can be collected at each level.
\clearpage
\begin{longtable}{|>{\raggedright}p{1.5cm}|>{\raggedright}p{2cm}|>{\raggedright}p{2.3cm}|>{\raggedright}p{2.3cm}|>{\raggedright}p{2.3cm}|>{\raggedright\arraybackslash}p{2.3cm}|}
\caption{MaaS data collection aggregated by level and stakeholder.}
\label{tab:maas_data_summary} \\
\hline
 &
  \textbf{Level 0 (No integration)} &
 \textbf{ Level 1  (Information Integration)} &
\textbf{ Level 2  (Integration of search \& reservation/payment)} &
 \textbf{Level 3  (Service Delivery Integration)} &
\textbf{ Level 4 (Policy integration)}  \\  \hline  \endhead
\textbf{Users }&
    No data collected by MaaS providers since there is no platform in place &
	User profile such as name, email, and phone number for registration and authentication purposes &
	User profile data including payment information such as credit card details for payment processing, location data for real-time traffic updates and directions, and trip data from multiple transportation providers to offer more comprehensive transportation services.\\ 
    User preferences and personal information for personalized services may also be collected. &
	Travel behavior and preferences to optimize transportation services.\\
    Advanced location data, such as real-time data from sensors and GPS, and data from external sources, such as weather and events, may also be collected to inform transportation planning.\\
    Data on transportation infrastructure, such as road conditions and public transit schedules, may also be collected to provide real-time information to users. &
	Sensitive personal information, such as biometric data, medical information, and travel history, to provide a complete end-to-end travel experience for users.
    Data from multiple transportation providers and infrastructure sources to offer a seamless and efficient travel experience.
    Analytics data, such as usage data and user behavior, to improve the MaaS platform and services. \\ \hline  
\textbf{Transport Providers} &
    Data collected directly by transport providers may include location data, trip data and passenger data to improve service offerings and route optimization &
	Basic information such as routes, schedules and fares for basic transportation options &
	More detailed information on routes, schedules and fares for multiple transportation options &
	  Location data, trip data, and passenger data for improved service offerings and route optimization. Real-time data from sensors and GPS for improved service delivery. &
	  Location data, trip data, and passenger data for improved service offerings and route optimization. Real-time data from sensors and GPS for improved service delivery. \\ \hline
	\textbf{City Authorities }&
	  Data collected directly by city authorities may include traffic data,  public transit data, and infrastructure data to inform transportation planning and policy &
	  No data collected &
	  Traffic data, public transit data, and infrastructure data to inform transportation planning and policy. &
	  Traffic data, public transit data, and infrastructure data to inform transportation planning and policy. &
	  Traffic data, public transit data, and infrastructure data to inform transportation planning and policy. Data on environmental impact and energy use to inform policy decisions. \\ \hline
	\textbf{Third-Party  Services} &
	  No data collected by MaaS providers since there is no platform in place &
	  No data collected &
	  Marketing data, user data to deliver targeted ads and promotions. Data on reservations  and payments to facilitate the transportation booking process. &
	  Analytics data, such as usage data and user behavior, to improve the MaaS platform and services. &
	  Marketing data, user data to deliver targeted ads and promotions. Data on policy compliance  to inform decision-making. \\ \hline
\end{longtable}

\section{Privacy Laws and Regulations in MaaS}
\label{privacy_laws_and_regulations_in_maaS}
Privacy rules are crucial in MaaS. This later makes heavy use of people's data for route optimization, pricing and quality of service. This highlights the importance of privacy laws and regulations for data collection and use, ensuring user trust  \cite{pollicino21}. This section presents these rules in MaaS, including some policies, their impact, and implications for service providers.

\subsection{Global Overview of Privacy Laws and Regulations in MaaS}
Privacy laws and rules are very important for MaaS. They create a framework to protect data and user privacy. One of the most important laws is the General Data Protection Regulation (GDPR), which started in Europe. This law has had a big impact on MaaS providers globally \cite{kuner20} \cite{bradford}. It sets out key principles including data minimization, obtaining explicit user consent before data collection, and offering users the 'right to be forgotten'. For example, a leading MaaS provider in Germany had to extensively revise its data collection procedures to ensure explicit user consent, signifying the profound influence of GDPR on the MaaS industry in Europe.

Simultaneously, in the United States, the California Consumer Privacy Act (CCPA) sets stringent rules for companies handling personal data. For example, the CCPA requires MaaS providers to disclose their data collection, selling, and sharing practices to users. It has led to substantial changes in the operations of MaaS providers, exemplified by a notable provider in California that had to significantly modify its data handling practices to ensure compliance with the CCPA \cite{preston19}.

Beyond these two significant regulations, other regional or national privacy laws also impact MaaS. For example, the Personal Information Protection and Electronic Documents Act (PIPEDA) in Canada \cite{pipeda20} is currently being re-evaluated (Bill C-27) to reflect the increasing obligations of companies handling personal data \cite{c27} . The Personal Data Protection Act (PDPA) in Singapore \cite{pdpa12} is also an important text dealing with the growing importance of privacy rules. While broadly following the principles of GDPR and CCPA, these regulations introduce unique elements. For instance, Bill C-27 emphasizes accountability, requiring organizations to appoint a data protection officer, while PDPA in Singapore stresses the protection of children's data.

Furthermore, PIPEDA governs how organizations collect, use, and disclose personal information in the course of commercial activities. To protect MaaS data under PIPEDA, users are encouraged to exercise their rights and be cautious about the personal information they share online. Privacy-enhancing tools like VPNs and encrypted messaging services are also recommended.

In Quebec, the Access to Information Commission issued a reminder in September 2021 about the enactment of certain provisions of Bill 25, the Act to modernize legislative provisions relating to the protection of personal information in the private sector. This reform adjusts the rules for protecting personal information in Quebec to address the challenges posed by the digital and technological environment. Bill 25 imposes additional obligations on private businesses and promises increased protection of personal information, new rights for citizens, and more responsible and transparent management of personal information. The changes will be phased in over three years, until 2024 \cite{bill25}.

\subsection{The Role of Privacy Impact Assessments (PIAs)}

Among the more demanding measures found in the aforementioned laws, several require, expressly or not, that companies draw up PIA to demonstrate the diligence employed in the management of personal data. PIAs are a fundamental tool within the global privacy regulatory landscape and have significant implications for how MaaS providers handle personal data. These assessments involve a systematic review of the potential effects of a project, policy, or system on privacy rights, paving the way for identifying privacy risks and implementing appropriate measures to mitigate them \cite{clarke09}. Though not universally legally mandated, PIAs are considered best practices in many jurisdictions. For example, the Office of the Privacy Commissioner of Canada (OPC) advocates for PIAs to determine how new or modified initiatives might impact personal privacy. In some provinces like British Columbia, PIAs are legally required for certain contexts \cite{georgiadis22}. The same is true of new laws passed in Quebec \cite{bill25}. 

Within the MaaS context, PIAs enable service providers to scrutinize and address the privacy implications of their offerings proactively. PIAs ensure the identification and mitigation of potential privacy risks before they materialize, thus creating a preventative approach to privacy protection. This can involve scrutinizing data collection, storage, and processing procedures, assessing potential impacts on user privacy, and instituting necessary safeguards.

Implementing PIAs also allows MaaS providers to demonstrate their dedication to privacy, fostering trust among users and aiding in maintaining regulatory compliance. Moreover, PIAs can guide the design and deployment of MaaS services, promoting the integration of privacy measures from the onset. This preventive and proactive strategy aligns with the "privacy by design" approach advocated by leading data protection frameworks like the GDPR \cite{cavoukian09}.

\subsection{Implications for MaaS Providers}

The far-reaching privacy laws and regulations impose a profound influence on the operations, business models, and consumer relations of MaaS providers \cite{quach22}. With the increasing digitization of services, MaaS providers rely heavily on user data to enhance their services and offer personalized experiences. As a result, these providers must navigate a complex landscape of laws that protect user privacy while ensuring their services remain competitive and user-centric \cite{maasalliance23}.

One of the primary impacts of these laws is on the operations of MaaS providers. Laws such as GDPR and CCPA mandate strict guidelines on data collection, usage, and storage. 

The data handling procedures of MaaS providers frequently need to undergo extensive adjustments in order to comply with these regulations. These adjustments may involve implementing systems to collect and manage user consent, creating tools for data minimization, and enhancing data security architecture \cite{maasalliance23}.

Additionally, these restrictions significantly impact the business models of MaaS providers. Laws on data collection and usage can force providers to rethink their strategies for data-driven monetization. The business models must balance the need for data to improve services with the imperative of user privacy. This balance requires careful planning and execution to ensure both profitability and compliance \cite{pagoni22}. This is especially true in light of the severe consequences for breaking the law, outlined in the GDPR, Bill C-27 in Canada, and L25 in Quebec. These laws also strongly affect how MaaS companies interact with their customers. Privacy laws often ask for openness about how data is used, which can help create trust with customers. By demonstrating their compliance with these laws, MaaS providers can portray themselves as responsible custodians of user data, which can help in fostering stronger consumer relations \cite{pangbourne20}.

Despite the potential benefits, compliance with these laws presents significant challenges for MaaS providers. The laws often differ across jurisdictions, which means that providers operating in multiple regions need to navigate a complex, varied landscape of privacy regulations. Moreover, as these laws evolve over time, maintaining compliance can be an ongoing challenge that requires constant vigilance and adaptability \cite{maasalliance23}.

Indeed, navigating the evolving landscape of privacy laws and regulations is a significant challenge for MaaS providers. However, this challenge also presents an opportunity for innovation and improvement in data privacy practices. In this context, the classification of data privacy methods emerges as a critical area of focus. By understanding and implementing these methods, MaaS providers can not only ensure compliance with laws but also offer a secure, privacy-respecting service to users. The subsequent section delves into this topic, providing a comprehensive classification of data privacy methods in MaaS.

\section{Classification of data privacy methods in MaaS}
\label{classification_of_data_privacy_methods_in_maas}

\begin{figure*}[!h]
\centering
\includegraphics[width=\linewidth,height=10cm]{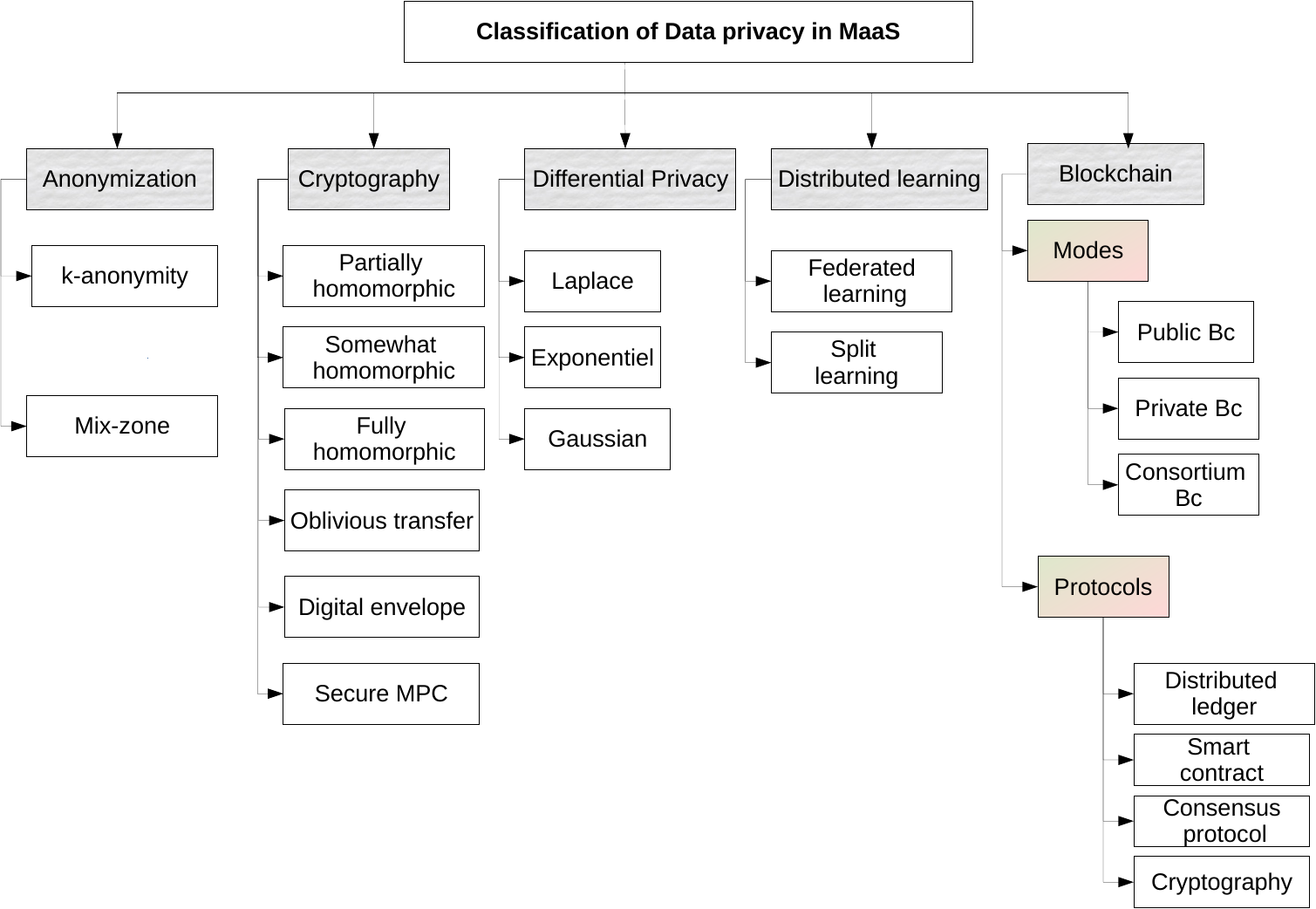}
\caption{Taxonomy of Privacy-preserving methods for MaaS}
\label{fig_1}
\end{figure*}

Given the pivotal role of data privacy in MaaS \cite{shokri11}, substantial research has been devoted to the development of various privacy-preserving techniques. These methods encompass anonymization, cryptography, differential privacy, distributed learning methods and blockchain. In the subsequent subsections, we will discuss each of these methods, provide an overview of their current state of the art, and compare their effectiveness in preserving privacy. Figure \ref{fig_1} depicts the proposed classification conducted by several journal/conference papers.

\subsection{Anonymization methods}

Within the realm of data anonymization, k-anonymity and mix zones stand out as two frequently utilized strategies \cite{pham17}, \cite{almomani18}, \cite{hongcheng21b}, \cite{martelli22}. 

The k-anonymity method is a privacy protection technique that ensures that a user's location cannot be distinguished from $k-1$ other users using generalization and suppression algorithms. To achieve this, a trusted third party (TTP) operates the location privacy server that anonymizes user locations. When a user sends a location query, the TTP calculates a set of $k$ users and provides an obfuscation area containing $k$ positions, including the querying user's location. This approach is useful for applications that do not require true or pseudo-identity, such as finding nearby restaurants or notifying a user of sale prices. However, it is not useful when identity information is necessary for the LBS to provide its services \cite{zhangb2019}.

K-anonymization has been extended in two ways. The first involves using multiple distributed servers or peer-to-peer communication instead of a single trusted anonymizer, while the second involves constraining the users included in $k$ based on contextual knowledge that an adversary may possess. P-sensitivity guarantees that the key attributes have at least $p$ different values within the $k$ user set, while l-diversity ensures that the user's location is unidentifiable from a set of $l$ different physical locations. Historical k-anonymity provides guarantees for moving objects.

Mix-zones are a privacy protection technique that does not require user identity information. It involves continually changing a user's pseudonym within a mix-zone to maintain privacy \cite{beresford04}. This technique has been applied in various applications, including location privacy in vehicular networks \cite{abdueli22} and mobile crowd sensing \cite{wang2020b}. 

The limitations of anonymization methods, discussed in \cite{domingo08}, arise from the compromise between privacy and data utility. Techniques like k-anonymity, p-sensitive k-anonymity, and l-diversity lack comprehensive privacy protection, while t-closeness sacrifices attribute correlations for complete privacy. Computational strategies for these methods often involve generalization and suppression, distorting numerical attributes into categorical ones. The absence of a clear procedure for t-closeness further complicates the issue. Mix-zones, similarly, encounter challenges in balancing anonymity and meaningful data representation.
 
Addressing the challenges in implementing k-anonymity and mix zones, as well as the call for more context-aware solutions, numerous recent studies have been undertaken. For instance, PrivateRide in \cite{pham17} addresses the issue of location privacy by providing user anonymity and hiding the origin and destination of the trip from the service provider (SP). Instead of sharing this sensitive information, an approximate origin and destination area are sent to the SP. The precise locations are only revealed to the driver once they are matched with a passenger. Al-Momani \cite{almomani18} proposed a privacy scheme in driverless cabs service (iRide) with analogy with PrivateRide. iRide relies on isolated execution environments; Intel SGX to anonymize pick-up and drop off locations. \cite{hongcheng21b} presented a protocol that employs RNE to calculate shortest distances effectively. Their approach utilizes hash functions that maintain certain properties, allowing the service provider (SP) to not only calculate distances between riders and drivers, but also select the nearest driver. Consequently, they eliminate the requirement for an additional cryptographic server. \cite{martelli22} proposes a privacy-enhancing technique based on the selection of points of interest (POIs) for pick-up and drop-off locations. This technique involve selecting POIs that are more crowded or more generic in order to obfuscate the exact location and identity the passengers. \cite{duan19} studied the matching performance loss in cloaking methods and propose to maximize the social welfare (minimize the overal pick-up distances) with pricing scheme to compensate loss of priority matching accuracy. 

\subsection{Cryptography or Private computation}

Cryptography has the potential to enhance data privacy in MaaS by protecting sensitive information, reducing the risk of data breaches, and preserving user privacy  \cite{luo19a}, \cite{luo22}, \cite{yu21b}, \cite{yu22}, \cite{yu20}, \cite{nabil21}, \cite{ulrich16}, \cite{fengwei18}, \cite{haining19}, \cite{ma22} . However, there are also some potential drawbacks to consider. For instance, private computation can require additional computational resources, which may slow down processing time and reduce system performance. Implementing private computation can also be complex and require expertise. 

Private computation's privacy guarantees may be constrained by factors such as the accuracy of data models and the quality of the data used. Furthermore, it may impede sharing data with third parties, which can be a disadvantage for MaaS applications reliant on data exchange. Finally, implementing private computation can be costly, particularly for small businesses or startups that may lack the resources to invest in privacy-preserving technologies \cite{ustundag22}. Among cryptography methods:

Partially homomorphic encryption allows for computation on only one type of operation (either addition or multiplication) while keeping the ciphertext encrypted \cite{ryu2023}. Somewhat homomorphic encryption enables computation on both addition and multiplication but has a limited number of operations that can be performed. Fully homomorphic encryption enables arbitrary computation on encrypted data \cite{lakhan22}.

Oblivious Transfer is a protocol in cryptography that enables a party to selectively receive one of several pieces of information from another party. The sender remains unaware of the recipient's selection, ensuring privacy. It is foundational to secure multi-party computations, playing a vital role in maintaining secrecy and trust in data exchange \cite{xianmin21}. 

A digital envelope is a cryptography concept that involves the use of two keys: a public key to encrypt data, and a private key to decrypt it. The term "digital envelope" is derived from the analogy of placing a sealed message (encrypted data) in an envelope for secure transmission.

Secure multiparty computation (MPC) is a technique that enables multiple parties to jointly compute a function over their private inputs without revealing them to each other \cite{nabil21}. It allows the parties to securely collaborate and compute a result that they would not be able to obtain individually due to the privacy of their inputs. 

The recent literature abounds with advancements in cryptographic methods, demonstrating their potential to deliver privacy-preserving solutions in MaaS applications. Luo et al. \cite{luo19a} constructed a privacy-preserving ride-matching scheme (pRide) with a light homomorphic encryption between rider's start point and driver's current location using road network distances and Yao's garble circuit. \cite{luo22} presents P$^2$Ride, a privacy-preserving ride-matching scheme for ridesharing services that uses homomorphic encryption and secure multi-party computation techniques to protect the privacy of passenger and driver locations while matching them efficiently. The system is evaluated using real-world datasets, and the authors claim that P$^2$Ride provides a practical solution that balances the efficiency and privacy concerns of both passengers and drivers. The article also discusses the importance of privacy in ridesharing services and highlights the potential privacy risks associated with existing ride-matching approaches. \cite{yu21b} proposed a zone-based minimum road travel time estimation technique with minimum travel time called PSride, using ciphertext packing protocols to schedule taxis with minimum additional travel time.  Yu et al. \cite{yu22} proposed an efficient exact shortest distance over encrypted location for privacy-preserving in an online ride-hailing. Based on SHE, \cite{yu20} proposed an encrypted aggregate distance computation method to calculate aggregate distances from a group of passengers to large-scale drivers. \cite{ulrich18} provides an user-privacy in ride-sharing to matches drivers with riders, a filter protocol based on homomorphic arithmetic secret sharing and secure two-party equality test to choose the drivers with whom the rider can travel. In \cite{nabil21} the authors present a protocol for ensuring user privacy in ridesharing services. The system utilizes secure multiparty computation techniques to compute the optimal route and fare for each ride without revealing sensitive user information. The proposed system allows for both transferable and non-transferable services, and it is evaluated through simulations and experiments to demonstrate its efficiency and effectiveness in preserving user privacy. The protocol also uses techniques such as pseudonymization, encryption, and anonymous authentication to protect user privacy.
 \cite{ulrich16} proposed a protocol that uses secure multiparty computation techniques to protect the privacy of users while determining the optimal meeting points for ridesharing. The protocol allows users to specify their preferences for the meeting point, such as a convenient location or a secure place, without revealing their exact location or identity. The proposed system is evaluated through simulations and experiments, and the results show that it can provide efficient and privacy-preserving meeting points for ridesharing services.   Fengwei et al. \cite{fengwei18} proposed an lightweight dynamic spatial query scheme named TRACE to protect the sensitive data for online ride-hailing providers and users based on random masking technique and point-in-polygon strategy. The scheme uses bilinear pairing for encrypting driver and rider locations, i.e, the encrypted space division information based on a quadtree structure and symmetric encryption which avoid disclosing business secrets and causing economic costs, then stored in rider and driver side. After a rider is paired with a driver. A secure communication channel is needed to negotiate a specific pick-up location. The integrity and authenticity of a message are validated using a digital signature. \cite{haining19} employed techniques such as road network embedding and homomorphic encryption to accurately and confidentially calculate the shortest distances between users in a ride-hailing system. Their approach was designed to be efficient and maintain the privacy of the users' information.
\cite{ma22} use somewhat homomorphic encryption and proposed a privacy-preserving cross-zone ride-matching scheme (CRide) for online ride-hailing while achieving an acceptable accuracy of matching algoritm.

\subsection{Differential Privacy}

Differential privacy reduces the risk of privacy breaches by adding a controlled amount of random noise to the data being analyzed. The idea is to add enough noise so that individual data points in the dataset cannot be linked to any specific individual with high confidence, while still allowing useful information to be extracted from the data \cite{xu20,khazbak18,belletti17,shen21}.

An algorithm is called epsilon-differentially private if it satisfies the definition of differential privacy with a parameter $\epsilon$ (epsilon). The parameter $\epsilon$ controls the amount of noise that is added to the output of the algorithm in order to protect privacy.

The definition of epsilon-differential privacy states that for any two datasets $D$ and $D'$, which differ by only one individual's record, and any output $S$ of the algorithm:

$Pr[S(D) \in A] \leq exp(\epsilon) * Pr[S(D') \in A]$

where $Pr[S(D) \in A]$ is the probability that the output $S$ of the algorithm on dataset D belongs to a set $A$, and $exp(\epsilon)$ is the privacy budget or privacy parameter. This definition states that the probability of obtaining a particular output from dataset $D$ is at most exp($\epsilon$) times the probability of obtaining the same output from dataset $D'$. This ensures that the algorithm does not reveal any information about any individual in the dataset with high probability.

In practice, the value of $\epsilon$ is chosen based on the sensitivity of the data being analyzed and the desired level of privacy protection. A smaller value of $\epsilon$ corresponds to stronger privacy protection but may result in more noise being added to the output, which could reduce the accuracy of the algorithm. Conversely, a larger value of $\epsilon$ corresponds to weaker privacy protection but may result in less noise being added to the output, which could improve the accuracy of the algorithm \cite{hsu14}.

Differential privacy methods use various probability distributions to add noise to queries and protect the privacy of sensitive data. Laplace distribution is a popular choice due to its heavy tail, which allows it to add more noise to extreme values. It is often used for queries with high sensitivity and limited privacy budget. Exponential distribution models the time between events in a Poisson process and is commonly used when the sensitivity of the query is unknown. Gaussian distribution is a bell-shaped distribution and is preferred when the sensitivity of the query is relatively small and a smaller amount of noise is required to maintain accuracy. The choice of distribution depends on the specific application and privacy requirements \cite{dwork2008}.

In the realm of applying differential privacy to ride-hailing and ride-sharing platforms, several approaches have been proposed, each with unique methodologies. For instance, Xu et al. \cite{xu20} proposed a geo-indistinguishability-based framework to perturb location in an online minimum bipartite matching problem and preserve the privacy of individuals on ride-sharing platforms while minimizing the total waiting time of passengers. The riders send perturbed location to an untrusted third-party platform. Then, this later performs an online assignment algorithm and establishes a direct connection with the requester.  The techniques used in \cite{shen21} include differential privacy and pseudonymization of the user's location data. The authors also suggest the use of a trusted third party to ensure the anonymity of the user's identity. \cite{belletti17} use differential privacy techniques to protect user privacy while collecting and analyzing data on user demand and mobility patterns for the optimization of fleet management. The system uses a hierarchical data aggregation technique to collect data at various levels of aggregation while ensuring privacy. The authors also propose a privacy-preserving algorithm for the optimization of fleet management that uses the collected data. \cite{khazbak18} and \cite{khazbak20} proposes a privacy-preserving framework for ride-hailing services that employs several privacy-preserving techniques, including location perturbation, randomized service selection, trajectory blurring, and differential privacy. These techniques work in unison to protect the user's location data and travel patterns while still enabling the service to deliver precise location-based services. Location perturbation obscures the user's actual location, randomized service selection limits the amount of information service providers can access, trajectory blurring obscures the user's travel path, and differential privacy adds an extra layer of protection for the user's location data. Together, these techniques ensure that ride-hailing service users' privacy is safeguarded without compromising the quality of the services provided.

\subsection{Distributed learning}

Distributed learning for data privacy refers to the use of distributed machine learning techniques to train models using data from multiple sources while preserving the privacy of each data source. Federated learning and split learning are two popular approaches to distributed learning for data privacy in MaaS \cite{duan22}.

Federated learning is a technique where multiple devices or edge servers train a shared model collaboratively without sharing raw data with a central server \cite{fuxun22}. Instead, each device performs local training on its own data, and only model updates are sent to the central server. The central server aggregates the updates and sends the updated model back to the devices, and the process repeats until convergence. Federated learning enables privacy preservation because raw data never leaves the devices, and the model updates are encrypted during transmission \cite{ustundag22}. The main drawback of federated learning is the need for a high-bandwidth and low-latency communication infrastructure to transmit the model updates efficiently. It can also suffer from communication overhead if the model is large or the number of devices is significant. Additionally, federated learning assumes that each device has an adequate amount of data to train on, which may not always be the case.

Split learning, on the other hand, is a technique where the model is split between a central server and multiple edge devices. Each device performs local training on a portion of the model and sends only the updated weights to the central server. The central server aggregates the updates and updates the model accordingly \cite{qiang22}. Split learning also enables privacy preservation because the raw data never leaves the edge devices, and only model weights are transmitted. Split learning requires a more complex model architecture, which may increase the computational cost of training. Additionally, split learning may not perform well when the edge devices have limited computing power or low-quality data. Furthermore, the edge devices may require frequent updates to their local models, which can increase the communication overhead.

\cite{yansheng22} introduces Fed-ltd, a cross-platform ride-hailing approach that uses federated learning to train a dispatching model while preserving user data privacy. The method partitions user data across platforms and trains a local model on each platform, which are then aggregated to create a global model. Real-world datasets are used to evaluate the approach, demonstrating superior performance compared to traditional methods.

\subsection{Blockchain}

A blockchain is a tamper-proof, transparent, decentralized and distributed ledger that stores transactions using cryptographic hashes. It prevents fraud and diminishes security threats. It consists of a chain of blocks where each block includes several immutable transactions inside the blockchain network. A block encompasses various information in two sections: The header (the block index, Merkle root, timestamp, block hash of the previous block, and the data themselves), the body that store the transaction details. The first block is called the genesis block or zero block which preserves the ownership of the transactions. Blockchain requires a public–private key pair for the validation of the data. Moreover, there are three types of blockchains, namely public (permissionless), private (permissioned controlled by the specific enterprise), and consortium (federated blockchain, similar to a private blockchain). Transaction in blockchain is managed using smart contracts under concrete conditions. In addition, BT encompasses several consensus algorithms, for instance, Proof-of-Work (PoW), Proof-of-Activity (PoA), Proof-of-State (PoS), Practical Byzantine Fault Tolerance (PBFT), Delegated PoS (DPoS), Proof of-Burn (PoB), Federated BFT (FBFT), and Proof-of-Elapsed Time (PoET), to ensure reliability within the network. The decentralized nature of the blockchain means that it operates on a peer-to-peer network, eliminating the need for intermediaries to verify and validate transactions. This makes it an ideal technology for various applications, including digital currencies, supply chain management, and identity management \cite{baza20}, \cite{baza21}.

Blockchain technology has the potential to enhance MaaS by improving the efficiency, security, and transparency of transportation and mobility systems \cite{prashant22}.

One of the main benefits of using blockchain in MaaS is the ability to securely and transparently store and track data related to mobility services, such as vehicle usage, payment information, and trip history. This allows for more efficient and effective management of mobility services, as well as improved trust and accountability between the various parties involved. The decentralized nature of blockchain can also help to eliminate intermediaries, reduce transaction costs, and increase the speed of transactions. This can make it easier and more cost-effective for users to access and use various mobility services, such as car-sharing, ride-hailing, and public transportation. Furthermore, the use of smart contracts can help to automate many of the processes involved in MaaS, such as payments, scheduling, and route planning. This can improve the overall efficiency and reliability of the mobility services, as well as increase user satisfaction.

\cite{meng19} primarily focuses on the use of blockchain technology to enable secure and decentralized record-keeping of ride-hailing transactions. The consortium blockchain-based system ensures that data related to ride-hailing transactions is stored in a tamper-proof and decentralized manner, reducing the risk of unauthorized access or manipulation of the data. Additionally, the paper proposes the use of vehicular fog computing, which involves using the computing resources of nearby vehicles to process and analyze data related to ride-hailing transactions. \cite{nguyen19} propose a permissioned blockchain-based approach for enhancing the security and privacy of user data in Mobility-as-a-Service (MaaS) transactions, which include ride-hailing, public transportation, and other transportation services. This is achieved through the use of blockchain technology, which enables secure and decentralized record-keeping of MaaS transactions, making it tamper-proof, secure and eliminating the need for commercial agreements between the various MaaS agents.. Additionally, the paper suggests the use of smart contracts, which are self-executing contracts that allow for automated and secure data sharing between different MaaS service providers, while still maintaining the privacy of user data. \cite{nesma22} adresses the evolution of ride-sharing services from centralized models to decentralized ones. The article discusses how decentralized ride-sharing services can enhance user privacy by reducing the amount of personal data shared with centralized authorities. In \cite{badr21}, a blockchain for real-time ride-sharing is presented with low communication and computation overheads. The ride share area is presented as multiple overlapping grids. The privacy of matching is achieved through encryption of offers and requests. The system employs encryption techniques to protect the sensitive data of users, such as their location and identity. Specifically, the system uses homomorphic encryption to encrypt the location of users in a way that enables ride matching without disclosing the exact location of users. The system also uses zero-knowledge proofs to enable users to prove their identity and eligibility to use the system without revealing their identity or personal information. \cite{baza20}, \cite{baza21} proposed a method to protect the privacy of ride-sharing participants using a public Blockchain system named B-Ride. The location of the rider is concealed through cloaking, a technique that divides the coverage area into large cells. The rider's location privacy is maintained by disclosing only the cell number of the driver or rider, rather than their exact location. Ride requests are uploaded to the Blockchain, and drivers locally match them before sending encrypted ride offers to the matched riders' public keys. However, the use of cloaking in B-Ride results in a privacy-accuracy tradeoff. While larger cells improve privacy, accuracy suffers as two distant users can be considered a match. In addition, B-Ride does not protect the driver's privacy. \cite{wang21} utilizes a consortium blockchain to protect ride-sharing services. To increase both transparency and security, an attribute-based proxy reencryption algorithm has been implemented within the system. \cite{kakkar22a} proposes a blockchain-based data pricing scheme called Block-CPS for car sharing over a 5G network, which is secure and efficient. An IPFS-based data storage system ensures secure and cost-efficient data transactions within Block-CPS. The article uses a non-cooperative game-theoretic approach to achieve optimal pricing for both vehicle owners and customers. Blockchain integration further enhances security. \cite{meng22} proposes a ride-hailing service that uses private smart contracts to provide anonymous and secure service. The smart contract mechanism ensures that user and driver identities are kept private and only authorized parties can access ride information. Payment is only made after the completion of the ride, and a reputation system incentivizes good behavior. The paper includes a prototype implementation and experimental evaluation to demonstrate the feasibility of the proposed system.

\subsection{Gaps and limitations in the current literature}
Table \ref{tab:table1} provides a summary of the key aspects covered in the studied papers organized by publication year, and various data privacy techniques and technologies relevant to MaaS. The goal of the Table \ref{tab:table1} is to provide a quick overview of the current literature on data privacy in MaaS, showing which papers have covered which MaaS services and privacy techniques. This allows for easy identification of the coverage and gaps in the existing literature.

Based on the current literature, several gaps and limitations can be identified in the area of data privacy within MaaS. Addressing these gaps and limitations  can contribute to a more comprehensive understanding of MaaS  data privacy and help in the development of more effective privacy-preserving solutions :

\begin{itemize}
\item Lack of comprehensive coverage: There is no single paper that extensively covers all MaaS services, such as ridesharing, ride-hailing an other services, with a specific focus on data privacy. This suggests the need for more research that holistically addresses privacy challenges and requirements across various MaaS services.

\item Insufficient focus on emerging technologies: The existing literature may not sufficiently address the role of emerging technologies, such as edge computing in enhancing data privacy in MaaS. Further research is needed to explore the potential of these technologies to improve privacy-preserving solutions in MaaS.

\item Limited exploration of interdisciplinary approaches: Many papers focus on specific data privacy techniques, while a more interdisciplinary approach that combines multiple techniques may lead to more robust privacy-preserving solutions. Research that investigates the integration of various methods, such as anonymization, cryptography, and differential privacy, could yield more comprehensive results.

\item Inadequate attention to regulatory compliance: Although some papers discuss legal and regulatory aspects of data privacy in MaaS, there is still a need for more research that analyzes the implications of local and international data protection laws and how MaaS providers can ensure compliance.

\item Narrow evaluation of privacy techniques: Current literature may not provide a thorough evaluation of the effectiveness of different privacy techniques in MaaS settings. Future research should include more extensive empirical studies and comparisons of privacy-preserving methods, taking into account various performance metrics and real-world scenarios.

\item Insufficient focus on user-centered design: Research on data privacy in MaaS could benefit from a stronger emphasis on user-centered design and user privacy preferences. Investigating user perceptions and expectations related to privacy in MaaS can help to design more effective and user-friendly privacy-preserving solutions.
\end{itemize}

\begin{table*}[!t]
\caption{Relevant papers}
\label{tab:table1}
\centering
 \resizebox{\textwidth}{!}{ 
\begin{tabular}{p{0.7cm}p{0.8cm}p{1cm}p{1cm}p{1.4cm}p{1.4cm}p{1.4cm}p{1.85cm}p{1.6cm}p{1.5cm}}
\hline
 Paper & Year  & Ride-sharing & Ride-hailing & Edge computing & Blockchain & Distributed learning &  Anonymization & Cryptography & Differential privacy  \\ \hline
 
\cite{ulrich18} & 2018 & \checkmark	& --  &	--	& \checkmark	&	-- & 	\checkmark &  \checkmark &  --	 \\
\hline

\cite{martelli22} & 2022& 	& \checkmark &	--	&--	&	-- & --	& \checkmark &  --	 \\

\hline

\cite{baza20} & 2020& \checkmark	& -- &	--	&	\checkmark &	-- & \checkmark	&  \checkmark & -- 	 \\

\hline

\cite{baza21} & 2021 & \checkmark 	&--  &	--	& \checkmark	& --	 & \checkmark	& \checkmark &  	-- \\
\hline

\cite{wang21} & 2021 & \checkmark	& -- &	--	& \checkmark	&--	 &-- 	& \checkmark &  --	 \\
\hline

\cite{khazbak18} & 2018 & --	& \checkmark &	--	&	-- &--	 & \checkmark 	& -- &  --	 \\
\hline

\cite{khazbak20} & 2020 & --	& \checkmark &	--	&	-- &	--  & \checkmark	& -- &  --	 \\
\hline

\cite{hongcheng21b} & 2021 & --	& \checkmark &	--	& --	&	-- & --	& \checkmark & \checkmark 	 \\
\hline

\cite{yu20} & 2020 & --	& \checkmark  &	--	&	-- &	-- & 	-- & \checkmark & -- 	 \\

\hline

\cite{yu21b} & 2021 & --	& \checkmark &	--	&--	&--	 & --	& \checkmark &  --	 \\

\hline

\cite{luo19b} & 2019 & --	& \checkmark &		--& --	&	-- & -- 	& \checkmark & -- 	 \\
\hline

\cite{luo19a} & 2019 & --	&  \checkmark &	--	& --	&	-- & --	& \checkmark &  --	 \\
\hline

\cite{yu19c} & 2019 & --	& \checkmark &	--	&	-- &	-- & --	& \checkmark &  --	 \\
\hline

\cite{duan19} & 2019 &-- 	&  \checkmark &	--	& --	&	-- & \checkmark	& -- &  --	 \\
 \hline
 
\cite{yu22} & 2022 & 	--&  \checkmark &	--	&--	&	-- & --	& \checkmark &  --	 \\
 
 \hline 
\cite{ulrich16} & 2016 & \checkmark	& --  &	--	&--	&--	 & --	& \checkmark &  --	 \\
  \hline

\cite{almomani18} & 2018 & --	&  \checkmark &	--	&--	&	-- & --	& -- &  \checkmark	 \\

\hline

\cite{bradford} & 2018 & --	&  \checkmark &	--	&--	&	-- & --	& -- &  \checkmark	 \\

\hline

\cite{ma22} & 2022 & --	& \checkmark  &	--	&	-- &	-- & --	& \checkmark &  --	 \\

\hline

\cite{meng22} & 2022 &-- 	& \checkmark  &	--	&	\checkmark & --	 &-- 	& \checkmark &  --	 \\

\hline

\cite{duan20} & 2020 & \checkmark	& -- &	--	&	-- &--	 & \checkmark	& -- &  --	 \\
 
\hline

\cite{nabil21} & 2021 & \checkmark	&  --&	--	&	-- &	-- & --	&  \checkmark &  --	 \\
\hline

\cite{fengwei18} & 2018 & --	& \checkmark &	--	&--	&	-- & --	&\checkmark & -- 	 \\
\hline

\cite{xu20} & 2020 & \checkmark	& -- &	--	&	--&	-- & --	& \checkmark &  	-- \\
\hline

\cite{shen21} & 2021 & --	& \checkmark &	\checkmark	&	--&	-- & \checkmark	& \checkmark &  --	 \\

\hline

%

\cite{meng19} & 2019 & --	& \checkmark &	\checkmark	&\checkmark	&	-- & --	& \checkmark &  --	 \\

\hline

\cite{xu20} & 2020 & \checkmark	&  --&	--	&--	&	-- & --	& -- &  \checkmark	 \\
 
\hline

\cite{badr21} & 2021 & \checkmark	&--  &		--&\checkmark	&	-- & \checkmark	& \checkmark &  --	 \\
\hline

\cite{luo22} & 2022 & \checkmark	& -- &	--	&	--&	-- & \checkmark	& \checkmark &  --	 \\
\hline

\cite{kakkar22a} & 2022 & \checkmark	& \checkmark &	--	&\checkmark&	-- & 	-- & \checkmark &  -- \\
\hline

\cite{yansheng22} & 2022 & --	& \checkmark &	--	&	--&	\checkmark & --	&-- &  	-- \\
\hline
 \end{tabular}}
\end{table*}

\section{Future Research Directions} 
\label{future_research_directions}

As MaaS data privacy continues to evolve, promising research directions are emerging. In this section, we will presents three possible areas of exploration: game theory, edge computing, and evaluation of data privacy methods in MaaS. The ongoing exploration of these research directions will contribute to design more robust and context-aware privacy-preserving mechanisms.

Game theory can be a useful tool in the context of data privacy as it can analyze the incentives with respect to the collection and use of personal data, as well as the trade-off between privacy and convenience. It can also be used to analyze the strategies that individuals and organizations adopt to protect their privacy and control their personal data. Deng et al. propose a new framework for designing incentives and profit-sharing mechanisms in multi-modal transportation networks that encourage efficient and sustainable transportation solutions \cite{deng22}. Zheng et al. propose a method for mobile location privacy access control using game theory to address privacy concerns that arise from the use of LBS \cite{zheng18}. This method sets weight coefficients for location privacy factors, calculates an access control threshold based on mobile location privacy leakage, and determines an access control strategy based on a comparison with the threshold and a selection from a strategy matrix.

Edge computing can also enhance data privacy in MaaS in several ways. Firstly, it allows for local data processing and aggregation, reducing the risk of data breaches and protecting individual user privacy \cite{lin20,shen21}. Second, it supports federated learning, enabling data analysis without transmitting data to a central server. Third, it provides trusted execution environments, ensuring secure processing of sensitive data. Fourth, it allows for user-controlled data storage and processing, reducing the amount of data transmitted and stored centrally. Additionally, edge computing provides real-time processing, improving efficiency and enhancing the user experience by enabling the implementation of sophisticated algorithms that can predict ride demand, optimize routes, and manage traffic \cite{lin20}. Future research can explore trade-offs between privacy, latency, and resource consumption in edge computing scenarios, leading to more efficient and scalable privacy-preserving solutions.

Comparing data privacy methods in MaaS can be challenging due to several factors. The lack of standardization makes it difficult to measure or evaluate privacy, leading to challenges in comparing different methods and establishing best practices. Privacy methods and techniques are context-dependent, meaning that what works well in one context may not be appropriate for another. Privacy measures often involve trade-offs between privacy protection and other factors such as usability, functionality, and cost. The data privacy landscape is constantly evolving, making it difficult to evaluate the effectiveness of privacy methods over time and keep up with the latest developments. Finally, limited data availability on privacy breaches or violations makes it challenging to assess the effectiveness of different privacy methods. Nevertheless, several methods can be used in MaaS, such as PIA, which can systematically evaluate potential privacy risks and impacts. Privacy metrics, such as the number of data breaches or requests for data deletion, can also be developed. User feedback can be collected through surveys or interviews to identify areas of concern and improvement. External auditing can assess compliance with regulations, and internal monitoring, such as access log monitoring and security audits, can ensure that data privacy measures are effective \cite{callegati16}.

\section{Conclusions} 
\label{conclusions}
Data privacy techniques are crucial in MaaS to protect users' sensitive data from privacy risks. MaaS manipulate a wide range of data types, including location data, personal information, health information, and biometric data, which must be handled with care to prevent malicious activities such as cybercrime, surveillance and discrimination.

Laws and regulations play an essential role in protecting users' privacy in MaaS, and MaaS applications should comply with these regulations to ensure that users' data is used ethically and transparently. The GDPR and CCPA are examples of regulations that have been enacted to ensure that users' privacy is protected.

However, implementing data privacy techniques in MaaS poses significant challenges, including the need to balance privacy and convenience, the complexity of data flows, and the need for interoperability and standardization across different systems.

Privacy techniques such as anonymization, cryptography, differential privacy, distributed learning methods and blockchain are emerging, which can enhance data privacy in MaaS while preserving data utility. Additionally, the use of decentralized and distributed architectures could enhance privacy and security while reducing the need for centralized data management, leading to more widespread adoption and usage of these services.

Looking towards the future, there is a need for further research in data privacy for MaaS.  Game theory can be utilized to identify privacy risks and design privacy-preserving mechanisms. Edge computing is a more secure and private way of processing and storing data, as it reduces the amount of data transmitted to centralized servers. Evaluating different privacy-enhancing technologies is necessary to determine their effectiveness in protecting user privacy while ensuring data quality and service efficiency. Advancing our understanding of data privacy in MaaS through research will lead to more trustworthy and secure systems, benefiting both users and service providers.

\section{Declaration of Generative AI and AIassisted technologies in the writing process}
During the preparation of this work, the authors utilized ChatGPT by OpenAI in order to assist with content correction and linguistic refinement. After using this tool, the authors reviewed and edited the content as necessary and take full responsibility for the content of the publication.

\bibliographystyle{elsarticle-num}
\bibliography{maas_references.bib}
\end{document}